## ASTRONOMY

# Einstein@Home discovers a radio-quiet gamma-ray millisecond pulsar

Colin J. Clark,[1,2,3]* Holger J. Pletsch,[1,2] Jason Wu,[4] Lucas Guillemot,[4,5,6] Matthew Kerr,[7]
Tyrel J. Johnson,[8†] Fernando Camilo,[9] David Salvetti,[10] David Anderson,[12]
Carsten Aulbert,[1,2] Christian Beer,[1,2] Oliver Bock,[1,2] Andres Cuéllar,[1,2] Heinz-Bernd Eggenstein,[1,2,13]
Henning Fehrmann,[1,2] Michael Kramer,[3,4] Shawn A. Kwang,[11] Bernd Machenschalk,[1,2] Lars Nieder,[1,2]
Markus Ackermann,[14] Marco Ajello,[15] Luca Baldini,[16] Jean Ballet,[17] Guido Barbiellini,[18,19]
Denis Bastieri,[20,21] Ronaldo Bellazzini,[22] Elisabetta Bissaldi,[23,24] Roger D. Blandford,[25]
Elliott D. Bloom,[25] Raffaella Bonino,[26,27] Eugenio Bottacini,[25] Terri J. Brandt,[28] Johan Bregeon,[29]
Philippe Bruel,[30] Rolf Buehler,[14] Toby H. Burnett,[31] Sara Buson,[28†] Rob A. Cameron,[25]
Regina Caputo,[32] Patrizia A. Caraveo,[10] Elisabetta Cavazzuti,[33] Claudia Cecchi,[34,35] Eric Charles,[25]
Alexandre Chekhtman,[8†] Stefano Ciprini,[34,36] Lynn R. Cominsky,[37] Denise Costantin,[21] Sara Cutini,[34,36]
Filippo D'Ammando,[38,39] Andrea De Luca,[10,40] Rachele Desiante,[26,41] Leonardo Di Venere,[23,24]
Mattia Di Mauro,[25] Niccolò Di Lalla,[16] Seth W. Digel,[25] Cecilia Favuzzi,[23,24] Elizabeth C. Ferrara,[28,42]
Anna Franckowiak,[14] Yasushi Fukazawa,[43] Stefan Funk,[44] Piergiorgio Fusco,[23,24] Fabio Gargano,[24]
Dario Gasparrini,[34,36] Nico Giglietto,[23,24] Francesco Giordano,[23,24] Marcello Giroletti,[38]
Germán A. Gomez-Vargas,[45,46] David Green,[28†] Isabelle A. Grenier,[17] Sylvain Guiriec,[28,48]
Alice K. Harding,[28] John W. Hewitt,[49] Deirdre Horan,[30] Guðlaugur Jóhannesson,[50,51] Shiki Kensei,[43]
Michael Kuss,[22] Giovanni La Mura,[21] Stefan Larsson,[52,53] Luca Latronico,[26] Jian Li,[54]
Francesco Longo,[18,19] Francesco Loparco,[23,24] Michael N. Lovellette,[7] Pasquale Lubrano,[34]
Jeffrey D. Magill,[47] Simone Maldera,[26] Alberto Manfreda,[16] Mario N. Mazziotta,[24] Julie E. McEnery,[28,47]
Peter F. Michelson,[25] Nestor Mirabal,[28†] Warit Mitthumsiri,[55] Tsunefumi Mizuno,[56]
Maria Elena Monzani,[25] Aldo Morselli,[46] Igor V. Moskalenko,[25] Eric Nuss,[29] Takashi Ohsugi,[56]
Nicola Omodei,[25] Monica Orienti,[38] Elena Orlando,[25] Michele Palatiello,[18,19] Vaidehi S. Paliya,[15]
Francesco de Palma,[24,57] David Paneque,[58] Jeremy S. Perkins,[28] Massimo Persic,[18,59]
Melissa Pesce-Rollins,[22] Troy A. Porter,[25] Giacomo Principe,[44] Silvia Rainò,[23,24] Riccardo Rando,[20,21]
Paul S. Ray,[7] Massimiliano Razzano,[16] Anita Reimer,[25,60] Olaf Reimer,[25,60] Roger W. Romani,[25]
Pablo M. Saz Parkinson,[61,62,63] Carmelo Sgrò,[22] Eric J. Siskind,[64] David A. Smith,[65] Francesca Spada,[22]
Gloria Spandre,[22] Paolo Spinelli,[23,24] Jana B. Thayer,[25] David J. Thompson,[28] Diego F. Torres,[54,66]
Eleonora Troja,[28,47] Giacomo Vianello,[25] Kent Wood,[67†] Matthew Wood[25]

Millisecond pulsars (MSPs) are old neutron stars that spin hundreds of times per second and appear to pulsate as their emission beams cross our line of sight. To date, radio pulsations have been detected from all rotation-powered MSPs. In an attempt to discover radio-quiet gamma-ray MSPs, we used the aggregated power from the computers of tens of thousands of volunteers participating in the Einstein@Home distributed computing project to search for pulsations from unidentified gamma-ray sources in Fermi Large Area Telescope data. This survey discovered two isolated MSPs, one of which is the only known rotation-powered MSP to remain undetected in radio observations. These gamma-ray MSPs were discovered in completely blind searches without prior constraints from other observations, raising hopes for detecting MSPs from a predicted Galactic bulge population.

## INTRODUCTION

Since its launch in 2008, the Large Area Telescope (LAT) on board the Fermi Gamma-ray Space Telescope (1) has detected pulsations from more than 200 gamma ray–emitting rotation-powered pulsars (see http://tinyurl.com/fermipulsars). Nearly half of these are millisecond pulsars (MSPs), old neutron stars thought to have been spun up by accreting matter from a companion star (2, 3), reaching fast rotation rates of hundreds of revolutions per second (4). The first gamma-ray MSPs were discovered shortly after Fermi's launch by folding gamma-ray photon arrival times using rotational ephemerides from concurrent radio telescope observations (5). Dedicated radio searches targeting unidentified LAT sources (6) have also led to the discovery of more than 80 new MSPs.

However, radio pulsations from many gamma-ray pulsars are not detectable from Earth because of their low intrinsic luminosity or their radio beams not intersecting our line of sight. These radio-quiet pulsars can only be found by directly searching for pulsations at unknown frequencies in gamma-ray data. These "blind" searches in LAT data (7, 8) have discovered 57 young pulsars, 53 of which are radio quiet (9–11). However, the blind search detection of gamma-ray MSPs is challenging. Only one gamma-ray MSP (12), also detected in radio (13), has been found in previous blind searches, and this was only possible with additional positional and orbital constraints from optical and x-ray observations (14). With this exception, all rotationally powered MSPs have been discovered through their radio pulsations, limiting knowledge of the population of MSPs to nearby bright pulsars with radio emission beamed toward Earth (15).

Thus, in contrast to the large number of radio-quiet young pulsars seen by the LAT, no radio-quiet gamma-ray MSP has been found. Some disparity between the numbers observed in each of these two groups is expected; MSPs have wider radio beams that are visible from a larger range of viewing angles, making radio-quiet MSPs less common (16). However, the apparent absence of this source class so far is more likely to be due to the inherent difficulty of detecting the signal from an MSP at an unknown location. At higher pulsation frequencies, higher precision is required for the position-dependent "barycentering" corrections applied to photon arrival times to account for the Doppler shift due to Fermi's motion through the solar system. The localization region of a typical unidentified gamma-ray source, limited by the LAT's angular resolution to a few arc minutes (17), is far larger than the arc second precision required to detect gamma-ray pulsations from MSPs. Hundreds of thousands of sky locations covering the source localization region must therefore be searched, incurring a large computational cost.

In a recent study (18), we improved gamma-ray pulsar blind search methods to reduce their computational cost and increase their sensitivity. These methods are based on an efficient "semicoherent" first stage, in which only photons arriving within a short time of one another (within a "lag window") are combined coherently. More sensitive, but more computationally expensive, follow-up stages are then used to pick out and refine weak candidate signals. The efficient search-grid spacings derived from our study (18) allowed us to use a longer lag window and, hence, achieve higher sensitivity than was possible in previous surveys.

Using these techniques, we performed a survey (10) on the distributed volunteer computing system Einstein@Home (19) to search for pulsations from 152 pulsar-like unidentified sources (which have curved spectra and low flux variability) from the Fermi-LAT Third Source Catalog (3FGL) (17). The survey searched 5.5 years of improved "Pass 8" Fermi-LAT data (20). We searched the full gamma-ray source localization region for each source and over a range of spin frequencies and spin-down rates broad enough to cover all known young pulsars and MSPs. This search volume for each source was split into hundreds of thousands of smaller regions, each of which could be searched by a typical personal computer within a few hours. These are then distributed among the computers of tens of thousands of volunteers located across the globe, which run the search when otherwise idle. Using more than 10,000 years of volunteered central processing unit (CPU) time, this gamma-ray pulsar survey has discovered 17 young pulsars (10, 11).

## RESULTS

Among the sources searched by the Einstein@Home survey were two of the three unassociated 3FGL sources with the highest detection significance, 3FGL J1035.7−6720 and 3FGL J1744.1−7619 (18). Both of these have high Galactic latitudes and pulsar-like spectra, marking them as promising MSP candidates (21). These sources were searched

[1]Albert-Einstein-Institut, Max-Planck-Institut für Gravitationsphysik, D-30167 Hannover, Germany. [2]Leibniz Universität Hannover, D-30167 Hannover, Germany. [3]Jodrell Bank Centre for Astrophysics, School of Physics and Astronomy, University of Manchester, Manchester M13 9PL, UK. [4]Max-Planck-Institut für Radioastronomie, Auf dem Hügel 69, D-53121 Bonn, Germany. [5]Laboratoire de Physique et Chimie de l'Environnement et de l'Espace–Université d'Orléans/CNRS, F-45071 Orléans Cedex 02, France. [6]Station de radioastronomie de Nançay, Observatoire de Paris, CNRS/INSU, F-18330 Nançay, France. [7]Space Science Division, Naval Research Laboratory, Washington, DC 20375–5352, USA. [8]College of Science, George Mason University, Fairfax, VA 22030, USA. [9]Square Kilometre Array South Africa, Pinelands 7405, South Africa. [10]Istituto Nazionale di Astrofisica (INAF)–Istituto di Astrofisica Spaziale e Fisica Cosmica Milano, via E. Bassini 15, I-20133 Milano, Italy. [11]Department of Physics, University of Wisconsin-Milwaukee, P.O. Box 413, Milwaukee, WI 53201, USA. [12]Space Sciences Laboratory, University of California at Berkeley, Berkeley, CA 94720, USA. [13]Albert-Einstein-Institut, Max-Planck-Institut für Gravitationsphysik, D-14476 Potsdam-Golm, Germany. [14]Deutsches Elektronen Synchrotron (DESY), D-15738 Zeuthen, Germany. [15]Department of Physics and Astronomy, Clemson University, Kinard Laboratory of Physics, Clemson, SC 29634–0978, USA. [16]Università di Pisa and Istituto Nazionale di Fisica Nucleare, Sezione di Pisa I-56127 Pisa, Italy. [17]Laboratoire AIM, CEA-IRFU/CNRS/Université Paris Diderot, Service d'Astrophysique, CEA Saclay, F-91191 Gif sur Yvette, France. [18]Istituto Nazionale di Fisica Nucleare, Sezione di Trieste, I-34127 Trieste, Italy. [19]Dipartimento di Fisica, Università di Trieste, I-34127 Trieste, Italy. [20]Istituto Nazionale di Fisica Nucleare, Sezione di Padova, I-35131 Padova, Italy. [21]Dipartimento di Fisica e Astronomia "G. Galilei," Università di Padova, I-35131 Padova, Italy. [22]Istituto Nazionale di Fisica Nucleare, Sezione di Pisa, I-56127 Pisa, Italy. [23]Dipartimento di Fisica "M. Merlin" dell'Università e del Politecnico di Bari, I-70126 Bari, Italy. [24]Istituto Nazionale di Fisica Nucleare, Sezione di Bari, I-70126 Bari, Italy. [25]W. W. Hansen Experimental Physics Laboratory, Kavli Institute for Particle Astrophysics and Cosmology, Department of Physics and SLAC National Accelerator Laboratory, Stanford University, Stanford, CA 94305, USA. [26]Istituto Nazionale di Fisica Nucleare, Sezione di Torino, I-10125 Torino, Italy. [27]Dipartimento di Fisica, Università degli Studi di Torino, I-10125 Torino, Italy. [28]NASA Goddard Space Flight Center, Greenbelt, MD 20771, USA. [29]Laboratoire Univers et Particules de Montpellier, Université Montpellier, CNRS/IN2P3, F-34095 Montpellier, France. [30]Laboratoire Leprince-Ringuet, École polytechnique, CNRS/IN2P3, F-91128 Palaiseau, France. [31]Department of Physics, University of Washington, Seattle, WA 98195–1560, USA. [32]Center for Research and Exploration in Space Science and Technology (CRESST) and NASA Goddard Space Flight Center, Greenbelt, MD 20771, USA. [33]Italian Space Agency, Via del Politecnico snc, 00133 Roma, Italy. [34]Istituto Nazionale di Fisica Nucleare, Sezione di Perugia, I-06123 Perugia, Italy. [35]Dipartimento di Fisica, Università degli Studi di Perugia, I-06123 Perugia, Italy. [36]Space Science Data Center–Agenzia Spaziale Italiana, Via del Politecnico snc, I-00133, Roma, Italy. [37]Department of Physics and Astronomy, Sonoma State University, Rohnert Park, CA 94928–3609, USA. [38]INAF Istituto di Radioastronomia, I-40129 Bologna, Italy. [39]Dipartimento di Astronomia, Università di Bologna, I-40127 Bologna, Italy. [40]Istituto Universitario di Studi Superiori (IUSS), I-27100 Pavia, Italy. [41]Università di Udine, I-33100 Udine, Italy. [42]Center for Research and Exploration in Space Sciences and Technology II, University of Maryland, College Park, MD 20742, USA. [43]Department of Physical Sciences, Hiroshima University, Higashi-Hiroshima, Hiroshima 739-8526, Japan. [44]Friedrich-Alexander-Universität Erlangen-Nürnberg, Erlangen Centre for Astroparticle Physics, Erwin-Rommel-Strasse 1, 91058 Erlangen, Germany. [45]Instituto de Astrofísica, Facultad de Física, Pontificia Universidad Católica de Chile, Casilla 306, Santiago 22, Chile. [46]Istituto Nazionale di Fisica Nucleare, Sezione di Roma "Tor Vergata," I-00133 Roma, Italy. [47]Department of Physics and Department of Astronomy, University of Maryland, College Park, MD 20742, USA. [48]Department of Physics, The George Washington University, 725 21st Street, NW, Washington, DC 20052, USA. [49]Department of Physics, University of North Florida, 1 UNF Drive, Jacksonville, FL 32224, USA. [50]Science Institute, University of Iceland, IS-107 Reykjavik, Iceland. [51]NORDITA, Roslagstullsbacken 23, 106 91 Stockholm, Sweden. [52]Department of Physics, KTH Royal Institute of Technology, AlbaNova, SE-106 91 Stockholm, Sweden. [53]The Oskar Klein Centre for Cosmoparticle Physics, AlbaNova, SE-106 91 Stockholm, Sweden. [54]Institute of Space Sciences (IEEC-CSIC), Campus UAB, Carrer de Magrans s/n, E-08193 Barcelona, Spain. [55]Department of Physics, Faculty of Science, Mahidol University, Bangkok 10400, Thailand. [56]Hiroshima Astrophysical Science Center, Hiroshima University, Higashi-Hiroshima, Hiroshima 739-8526, Japan. [57]Università Telematica Pegaso, Piazza Trieste e Trento, 48, I-80132 Napoli, Italy. [58]Max-Planck-Institut für Physik, D-80805 München, Germany. [59]Osservatorio Astronomico di Trieste, Istituto Nazionale di Astrofisica, I-34143 Trieste, Italy. [60]Institut für Astro- und Teilchenphysik and Institut für Theoretische Physik, Leopold-Franzens-Universität Innsbruck, A-6020 Innsbruck, Austria. [61]Santa Cruz Institute for Particle Physics, Department of Physics and Department of Astronomy and Astrophysics, University of California at Santa Cruz, Santa Cruz, CA 95064, USA. [62]Department of Physics, The University of Hong Kong, Pokfulam Road, Hong Kong, China. [63]Laboratory for Space Research, The University of Hong Kong, Hong Kong, China. [64]NYCB Real-Time Computing Inc., Lattingtown, NY 11560–1025, USA. [65]Centre d'Études Nucléaires de Bordeaux Gradignan, IN2P3/CNRS, Université Bordeaux 1, BP120, F-33175 Gradignan Cedex, France. [66]Institució Catalana de Recerca i Estudis Avançats (ICREA), E-08010 Barcelona, Spain. [67]Praxis Inc., Alexandria, VA 22303, USA.

*Corresponding author. Email: colin.clark-2@manchester.ac.uk
†NASA Postdoctoral Program Fellow, USA.
‡Resident at Naval Research Laboratory, Washington, DC 20375, USA.

9 and 10 times, respectively, in a survey of unidentified Fermi-LAT sources performed with the Parkes radio telescope (6), but no pulsations were detected. Previous blind gamma-ray searches of these sources, including an earlier Einstein@Home survey (22) using less data and a shorter lag window, were also unsuccessful.

The latest Einstein@Home survey revealed gamma-ray pulsations from these sources, at frequencies of $\nu = 348$ Hz and $\nu = 213$ Hz, identifying them as isolated MSPs now known as PSR J1035−6720 and PSR J1744−7619. The gamma-ray pulsations had H test [a significance test for pulsations in unbinned event data (23)] values of $H = 389$ and $H = 494$, corresponding to single-trial false-alarm probabilities of $< 5 \times 10^{-68}$ and $< 4 \times 10^{-86}$, respectively. For comparison, Einstein@Home can perform fewer than $10^{20}$ independent trials in 1 year.

Following their discoveries, we refined the rotational and astrometric parameters of each pulsar by timing their gamma-ray pulsations. The resulting estimates and uncertainties on their timing and spectral parameters are given in Table 1, and the resulting integrated pulse profiles are shown in Fig. 1.

These MSPs remained undetected in more sensitive analyses of the earlier Parkes survey data (6), using the now-known pulsation periods. However, a weak signal [with average flux density $S_{1400} \approx 40 \, \mu$Jy (microjanskys) at a radio frequency of 1400 MHz] was detected from PSR J1035−6720 in four follow-up Parkes searches. Its estimated distance derived from its dispersion measure (DM = $84.16 \pm 0.22$ pc cm$^{-3}$) is $d \approx 1.46$ kpc or $d \approx 2.24$ kpc, according to the YMW16 (24) and NE2001 (25) Galactic electron-density models, respectively. PSR J1035−6720 therefore has a pseudo-luminosity of $L_{1400} = S_{1400} d^2 \sim 0.085$ to $0.20$ mJy (millijanskys) kpc$^2$. This is less luminous than the majority of known radio MSPs with measured pseudo-luminosities in the Australia Telescope National Facility (ATNF) Catalogue (26), as shown in Fig. 2.

In contrast, PSR J1744−7619 has remained undetected in two dedicated 3-hour follow-up searches on 19 March 2017 and 10 April 2017. These give an upper limit of $\sim 23 \, \mu$Jy on its radio flux density, below the $30 \, \mu$Jy threshold used by Abdo et al. (9) to classify gamma-ray pulsars as radio quiet. This threshold is chosen to represent a limiting sensitivity for pulsar surveys; a new generation of radio telescopes will be needed to discover new Galactic field pulsars with flux densities below this level (27).

In the absence of a dispersion measurement, the distance to PSR J1744−7619 is hard to constrain. A weak upper limit of $d \sim 1$ kpc can be obtained by assuming that its loss in rotational energy as it spins down is converted entirely into the observed gamma-ray emission. This distance limit implies a maximum $L_{1400} \approx 0.023$ mJy kpc$^2$. Only two known MSPs have lower reliably measured pseudo-luminosities (28, 29). Even if the lack of detection in these follow-up searches was due to temporary interstellar scintillation, the upper limit from the multiple shorter observations from the previous survey suggests a maximum $L_{1400} \approx 0.15$ mJy kpc$^2$, weaker than 90% of radio MSPs. PSR J1744−7619 therefore has (at least) an unusually low radio luminosity.

The fields of PSRs J1035−6720 and J1744−7619 were observed in x-rays by XMM-Newton as part of a complementary effort to discover radio-quiet MSPs by searching for possible x-ray counterparts to unidentified LAT sources (21). Positionally coincident x-ray counterparts were detected for both pulsars. Their gamma-ray–to–x-ray flux ratios ($F_\gamma/F_X \approx 700$ and 1100, respectively) are consistent with those of other known isolated gamma-ray MSPs (30). The gamma-ray spectral properties of these pulsars are also consistent with those of the MSPs studied in the Second Fermi-LAT Catalog of Gamma-ray Pulsars (*9*). Their x-ray and gamma-ray spectra therefore give us no reason to suspect that these radio-faint MSPs belong to another class of objects different from that of other gamma-ray emitting MSPs.

**Table 1. Measured and derived properties of the newly discovered MSPs.** For timing parameters, we report the mean values from a Monte Carlo timing analysis and 1σ uncertainties on the final digits in parentheses. The first set of uncertainties on spectral parameters is statistical and the second set estimates the effects of systematic uncertainties in the LAT's effective collecting area and the Galactic diffuse emission model. Timing parameters are in barycentric dynamical time (TDB) units. The spin-down power of PSR J1035−6720 has been corrected for the Doppler shift induced by its proper motion; only an upper limit is given for PSR J1744−7619, because this correction cannot be applied as a result of its uncertain distance. mas, milli–arc second.

| Parameter | PSR J1035−6720 | PSR J1744−7619 |
|---|---|---|
| **Timing parameters** | | |
| Reference time (MJD) | 55716 | |
| Data span (MJD) | 54682–57828 | |
| Right ascension (R.A.) (J2000.0), α (hh:mm:ss) | 10:35:27.478(1) | 17:44:00.488(2) |
| Declination (Decl.) (J2000.0), δ (dd:mm:ss) | −67:20:12.692(6) | −76:19:14.710(9) |
| Proper motion in R.A., $\mu_\alpha \cos \delta$ (mas year$^{-1}$) | −12(3) | −21(3) |
| Proper motion in Decl. $\mu_\delta$ (mas year$^{-1}$) | 1(3) | −7(3) |
| Spin frequency, ν (Hz) | 348.18864014054(8) | 213.33223675351(5) |
| Spin-down rate, $-\dot{\nu}$ (10$^{-15}$ Hz s$^{-1}$) | 5.633(1) | 0.4405(8) |
| Second frequency derivative, $|\ddot{\nu}|$ (10$^{-25}$ Hz s$^{-2}$) | < 1.1 | < 0.7 |
| Spin period, P (ms) | 2.8720063916972(7) | 4.687524094895(1) |
| Period derivative, $\dot{P}$ (10$^{-20}$ s s$^{-1}$) | 4.647(1) | 0.968(2) |
| **Derived parameters** | | |
| Galactic longitude, l (°) | 290.37 | 317.11 |
| Galactic latitude, b (°) | −7.84 | −22.46 |
| Spin-down power, $\dot{E}$ (10$^{33}$ erg s$^{-1}$) | 75.0 | < 3.7 |
| Characteristic age, $\tau_c$ (10$^9$ years) | 1.0 | 7.7 |
| Surface magnetic field, $B_S$ (10$^8$ G) | 3.7 | 2.2 |
| Light-cylinder magnetic field, $B_{LC}$ (10$^5$ G) | 1.4 | 0.2 |
| **Phase-averaged gamma-ray spectral parameters above 100 MeV** | | |
| Test statistic, TS | 1839.2 | 2492.2 |
| Photon index, Γ | 1.46 ± 0.07 ± 0.05 | 1.07 ± 0.10 ± 0.02 |
| Cutoff energy, $E_c$ (GeV) | 2.76 ± 0.26 ± 0.36 | 1.82 ± 0.19 ± 0.01 |
| Photon flux (10$^{-9}$ cm$^{-2}$ s$^{-1}$) | 24.4 ± 1.7 ± 1.5 | 19.2 ± 1.5 ± 1.0 |
| Energy flux, $F_\gamma$ (10$^{-12}$ erg cm$^{-2}$ s$^{-1}$) | 21.5 ± 0.8 ± 1.1 | 20.8 ± 0.8 ± 1.1 |
| **Off-pulse spectral parameters above 100 MeV** | | |
| Test statistic, TS | 7.4 | 33.8 |
| TS of exponential cutoff, TS$_{cut}$ | — | 8.3 |
| Photon index, Γ | — | 1.35 ± 0.70 ± 0.09 |
| Cutoff energy, $E_c$ (GeV) | — | 1.06 ± 0.76 ± 0.13 |
| Photon flux (10$^{-9}$ cm$^{-2}$ s$^{-1}$) | — | 1.9 ± 0.9 ± 0.1 |
| Energy flux, $F_\gamma$ (10$^{-12}$ erg cm$^{-2}$ s$^{-1}$) | — | 1.2 ± 0.3 ± 0.1 |

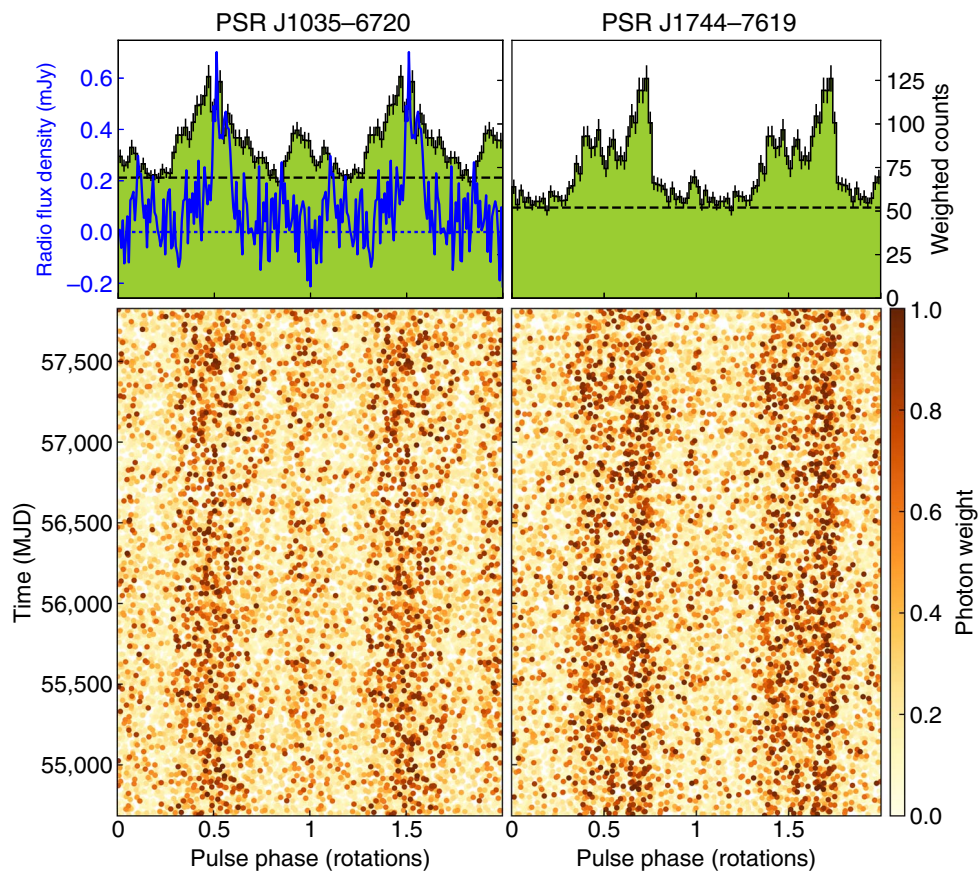

**Fig. 1. Pulsed signals from the two newly discovered MSPs.** Rotational phases of individual gamma-ray photons (**bottom**) and integrated pulse profiles (**top**) of the newly detected MSPs. Each photon has been assigned a weight, determined by its energy and arrival direction, representing the probability of it having come from the gamma-ray source in question. These weights are indicated by the color bar. In the top panels, the black dashed line shows the estimated background level, derived from the photon weights as in study of Abdo *et al.* (*9*). For PSR J1035−6720, the blue line shows the radio profile as measured by the Parkes radio telescope at 1400 MHz, averaged over four observations after subtracting an estimated baseline (dotted line). Two identical rotations are shown for clarity. MJD, Modified Julian Date.

One explanation for their weak radio pulsations could therefore be that their radio beams either clip our line of sight or, in the case of PSR J1744−7619, do not cross it at all. To investigate this scenario, we modeled the gamma-ray emission geometries by fitting simulated pulse profiles to the observed gamma-ray photon phases, using the fitting technique and models described in the study of Johnson *et al.* (*31*) and references therein. We considered three emission models: an outer gap (OG) model, a two-pole caustic (TPC) model, and a pair-starved polar cap (PSPC) model.

For PSR J1035−6720, only the PSPC model is consistent with the detection of radio pulsations. This model predicts that the gamma-ray pulse should lead the radio pulse, an unusual characteristic shared by only six other MSPs studied by Johnson *et al.* (*31*). The observed lag between the radio and gamma-ray pulses, shown in Fig. 1, is consistent with this prediction. However, the uncertainty of DM leads to an additional uncertainty of the arrival phase of the radio pulse of 0.15 rotations, larger than the observed lag between the two pulses. Because of this phase uncertainty and the faintness of the radio pulse, we have not attempted a joint fit of the radio and gamma-ray data.

The best-fitting models (TPC and PSPC) for PSR J1744−7619 have our line of sight cutting across the predicted radio emission cone and not merely clipping it. The lack of radio emission from PSR J1744−7619 is therefore not easily explained by small offsets of a few degrees from the best-fitting geometry. The best-fitting TPC model for this pulsar is further supported by a spectral analysis of off-pulse flux from PSR J1744−7619, which revealed likely magnetospheric unpulsed emission. Therefore, either the nondetection of radio pulsations is due to an extremely low intrinsic radio luminosity or it is at odds with the best-fitting emission models. More elaborate pulsar emission models, including emission from the current sheet (*32, 33*), may be required to address this tension.

## DISCUSSION

At least 12 high-confidence pulsar-like 3FGL sources at high Galactic latitudes have been identified but have remained undetected in repeated radio pulsation searches. Several convincing MSP candidates have also been highlighted by Saz Parkinson *et al.* (*21*). The most natural explanation for these sources is that they are either radio-quiet MSPs or MSPs in eclipsing binary systems, which are also difficult to detect in radio searches. Our results demonstrate that blind gamma-ray pulsation searches of LAT data are now able to find isolated MSPs among these sources without requiring any additional constraints from multiwavelength observations and therefore provide an entirely new method through which to investigate the Galactic MSP population.

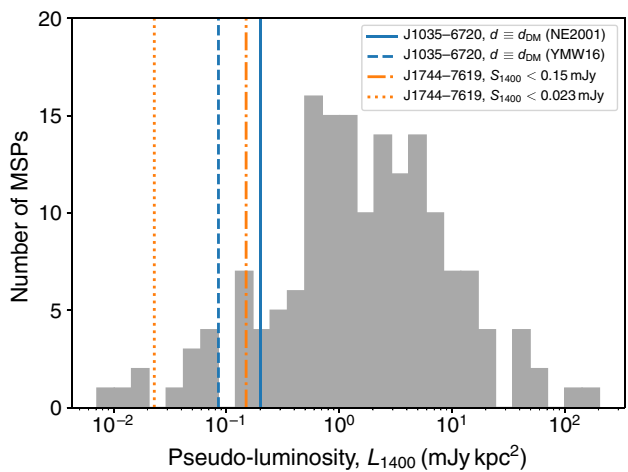

**Fig. 2. Pseudo-luminosities ($L_{1400} = S_{1400} \, d^2$) of known MSPs (including both Galactic field MSPs and those found in globular clusters) with flux density and distance measurements in the ATNF Catalogue (26), version 1.57.** The measured pseudo-luminosities of PSR J1035−6720 according to the NE2001 and YMW16 Galactic dispersion models are shown by solid and dashed vertical lines, respectively. The dotted and dashed-dotted vertical lines show the upper limits on the pseudo-luminosity of PSR J1744−7619, assuming a maximum distance of 1 kpc, from two dedicated follow-up radio observations and (a conservative estimate) from the 10 shorter observations of Camilo et al. (6), respectively.

In particular, blind gamma-ray searches may partially address the severe biases in the observed radio MSP luminosity distribution because of the difficulty of detecting faint MSPs (15). A sizeable sample of gamma ray–selected MSPs would also provide an unbiased estimate of the relative numbers of radio-loud and radio-quiet gamma-ray MSPs, constraining the ratio of the radio and gamma-ray beam solid angles, an important discriminator for pulsar emission models (16, 34). These models suggest that a large fraction of gamma-ray MSPs could be radio quiet; the contrast with the present situation of just three gamma ray–selected MSPs compared to almost 100 radio-loud gamma-ray MSPs illustrates both the technical barrier faced by blind searches for gamma-ray MSPs until now and their large discovery potential.

Furthermore, the rotational stability and low flux variability of MSPs mean that their recoverable signals will accumulate steadily over time. Blind gamma-ray searches will therefore become sensitive to fainter and more distant MSPs as the Fermi mission continues. The rotation rates of MSPs are also unaffected by glitches or timing noise, which often interrupt the signals from young pulsars. We may therefore expect future blind gamma-ray searches to find a larger proportion of MSPs compared to previous surveys, whose discoveries have been dominated by young pulsars.

The ability to detect radio-quiet MSPs in blind gamma-ray searches is also an important tool for the search for gamma-ray emission from dark matter. A number of unidentified LAT sources at high Galactic latitudes have curved spectra that are consistent with the annihilation of a candidate for dark matter in ultra-faint dwarf spheroidal galaxies or dark matter subhalos (35, 36). However, these sources are perhaps more likely to be MSPs, whose spectra are similar. If these sources continue to remain undetected at other wavelengths, then blind gamma-ray searches are necessary to confirm or rule out at least an isolated, radio-quiet MSP explanation. The observed GeV (billion electron volts) excess toward the Galactic center (37), another possible dark matter annihilation signal, is also consistent with an unresolved population of thousands of MSPs in the Galactic bulge (38–40). No more than a handful of these are expected to be detected with current radio telescopes because of the large distance to the bulge and the strong scattering in that region of the Galaxy (40). This latter issue does not affect gamma-ray pulsations, meaning that gamma-ray searches perhaps have the highest potential to uncover the brightest members of this population.

## MATERIALS AND METHODS
### LAT data preparation
The data in which the two MSPs were first detected covered approximately 5.5 years of LAT observations between 4 August 2008 and 6 April 2014. They were produced using preliminary internal versions of the Pass 8 instrument response functions (IRFs) and background models. Using gtselect, which is part of the Fermi Science Tools (http://fermi.gsfc.nasa.gov/ssc/data/analysis/software/), we selected SOURCE-class photons, within an 8° region of interest (ROI) around the positions of the two MSPs, with energies > 100 MeV and zenith angles < 100°. We excluded photons recorded when the LAT's rocking angle was > 52° and when the LAT was not operating in normal science mode.

A photon-weighting scheme developed by Kerr (41) greatly increases the sensitivity when searching for gamma-ray pulsations in gtselect. Using gtselect, which is part of the Fermi Science Tools (http://fermi.gsfc.nasa.gov/ssc/data/analysis/software/), we selected SOURCE-class photons, within an 8° region of interest (ROI) around the positions of the two MSPs, with energies > 100 MeV and zenith angles < 100°. We first carried out a spectral analysis of the target sources. The initial source model consisted of all sources in the 3FGL catalog within 13° of the target source and Galactic diffuse emission and isotropic diffuse background models. With this source model, we performed a binned maximum likelihood analysis using the sourcelike package (42). The spectrum of the target source was modeled with an exponentially cutoff power law (PL) with pre-factor $N_0$, spectral index $\Gamma$, and cutoff energy $E_c$:

$$\frac{dN}{dE} = N_0 \left( \frac{E}{1 \, \text{GeV}} \right)^{-\Gamma} \exp\left( -\left( \frac{E}{E_c} \right) \right) \tag{1}$$

The spectral parameters of the target and all sources within 5° were free to vary in the fit, as were the normalizations of the background models. From the results of this spectral fitting, probability weights were calculated for each photon using gtsrcprob.

After discovering the pulsars, we refined their spin parameters (see the "Pulsar parameter estimation" section) using extended data sets, which included LAT photons recorded from 4 August 2008 to 16 March 2017. These data consisted of publicly available Pass 8 R2 SOURCE-class photons, processed with the P8R2_SOURCE_V6 IRFs. When producing the extended data sets, we extended the size of the ROI to 15° and reduced the zenith angle cutoff to < 90°. To produce a source model for use with gtsrcprob, another binned spectral analysis was performed using gtlike, including sources within 20° of the pulsars, and with the latest gll_iem_v06.fits map cube (43) and iso_P8R2_SOURCE_V6_v06.txt template (http://fermi.gsfc.nasa.gov/ssc/data/access/lat/BackgroundModels.html) used to model the diffuse

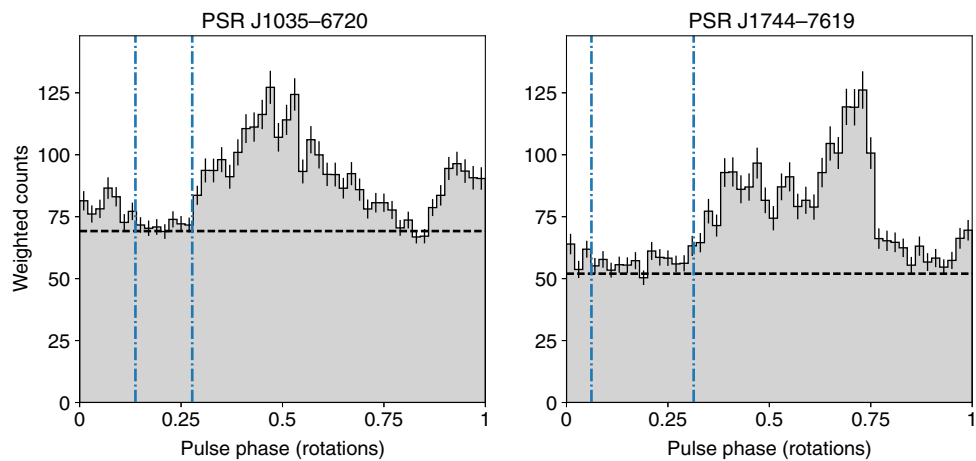

**Fig. 3. Pulse profiles of the two new gamma-ray MSPs, showing the phase definitions of the off-pulse regions (blue dashed-dotted lines) used to search for unpulsed gamma-ray emission.** The estimated background levels are shown by the black dashed lines. No significant off-pulse emission was detected from PSR J1035−6720. Off-pulse emission was detected with a significance of ~ 5σ from PSR J1744−7619.

emission. The source positions were fixed at the preliminary timing positions of the pulsars.

The estimates of the spectral parameters contain systematic uncertainties due to our choice of IRFs and diffuse background models. To estimate these systematic uncertainties, we performed the same spectral analysis with rescaled effective areas (http://fermi.gsfc.nasa.gov/ssc/data/analysis/scitools/Aeff_Systematics.html) and with the normalization of the Galactic diffuse emission rescaled to ± 6 % of the best-fit value.

The results of this spectral analysis for both pulsars are summarized in Table 1. The spectral properties ($\Gamma$ and $E_c$) of PSRs J1035−6720 and J1744−7619 are similar to those of the MSP population seen in the Second Fermi-LAT Catalog of Gamma-ray Pulsars (9).

## LAT off-pulse analysis

The integrated pulse profile for PSR J1744−7619 includes unpulsed emission at all phases above the estimated background level. We performed a spectral analysis of this "off-pulse" emission to determine whether this is likely to be magnetospheric emission or contamination from a nearby source. We performed a similar analysis to search for unpulsed emission from PSR J1035−6720.

We defined off-pulse regions for the two pulsars using the Bayesian block decomposition method described by Scargle et al. (44). The definitions of the off-pulse regions are shown in Fig. 3. We first computed residual test statistic maps (TS = 2 Δ log L, where Δ log L is the difference in log-likelihood between models with and without a putative source) with gamma-ray photons coming solely from the off-pulse region to look for putative sources around the pulsar position. No significant off-pulse emission was detected from PSR J1035−6720. This could be due to the low photon statistics in the defined off-pulse region; because the duty cycle for PSR J1035−6720 is close to 100%, a phase interval covering only ~14% of a rotation was selected for the analysis. Off-pulse emission at the position of PSR J1744−7619 was detected with TS = 33.8. Under the null hypothesis, the TS is approximately $\chi^2$ distributed with three degrees of freedom (from the normalization, spectral index, and cutoff that are free in this test). This TS therefore corresponds to a significance of ~ 5.0 σ. We computed the TS value with both PL and exponentially cutoff PL (PLEC) models to test for curvature of the gamma-ray spectrum. The off-pulse emission from PSR J1744−7619 shows marginal evidence of a cutoff, with $TS_{cut} = TS_{PLEC} − TS_{PL} = 8.3$ and one degree of freedom, corresponding to a significance of ~ 2.6σ. The positional consistency between the off-pulse source and the pulsar and the hint of a spectral cutoff suggest a magnetospheric origin for the unpulsed emission (9).

## Pulsar parameter estimation

After discovering the pulsars, we performed dedicated timing analyses using the extended data sets described above. To speed up our parameter estimation procedures, we applied a photon probability weight cutoff to remove the lowest-weighted photons. The cutoff values for each pulsar were chosen such that only 1% of the pulsation signal-to-noise ratio would be lost by this but around 90% of the lowest-weight photons would be removed. The analyses followed the timing procedure described by Clark et al. (10), with photon arrival times being corrected to the solar system barycenter using the JPL DE405 solar system ephemeris.

Starting with the signal parameters detected by the search (sky position, rotational frequency, and spin-down rate), we "phase-folded" the photon arrival times. We then fit template pulse profiles by minimizing the Bayesian information criterion (BIC) (45), the sum of the negative log-likelihood and a penalty factor proportional to the number of model parameters that prevents overfitting. We then varied the signal parameters using a parallelized affine-invariant Monte Carlo sampling algorithm (46, 47) to maximize the likelihood given our template pulse profile. This process continued iteratively until the likelihood stopped increasing: After each Monte Carlo step, we refolded the data with the most likely set of signal parameters, constructed a new template pulse profile, and repeated the Monte Carlo sampling.

We then added additional parameters to our phase model one by one, performed a further Monte Carlo step, and kept the new parameter in our model if its inclusion led to a decrease in the BIC. By this criterion, we were unable to measure a second frequency derivative or parallax from either pulsar. The inclusion of proper motion in our timing models led to a moderate decrease in the BIC (ΔBIC = −7.8) for PSR J1035−6720 and a significant decrease (ΔBIC = −40.6) for PSR J1744−7619.

Because we measured proper motion from each pulsar, a portion of the observed spin-down rate for each must be caused by the Doppler shift introduced by their velocity gaining an increasing radial component (9, 48), known as the Shklovskii effect. Correcting the observed spin-down rates (ν̇$_{OBS}$) of the pulsars for the Doppler-induced apparent spin-down (ν̇$_{SHK}$) due to their proper motion to retrieve their intrinsic spin-down rate (ν̇$_{INT}$) requires knowledge of their distances, which are uncertain without additional dispersion or parallax measurements. However, we can use the observed gamma-ray fluxes ($F_γ$) and proper motions (μ), and assume certain gamma-ray efficiencies (η), to retrieve constraints on the distances (*d*) by solving

$$\dot{ν}_{OBS} = \dot{ν}_{INT} + \dot{ν}_{SHK} = -\frac{F_γ f_Ω d^2}{η π I v} - \frac{μ^2 v d}{c} \qquad (2)$$

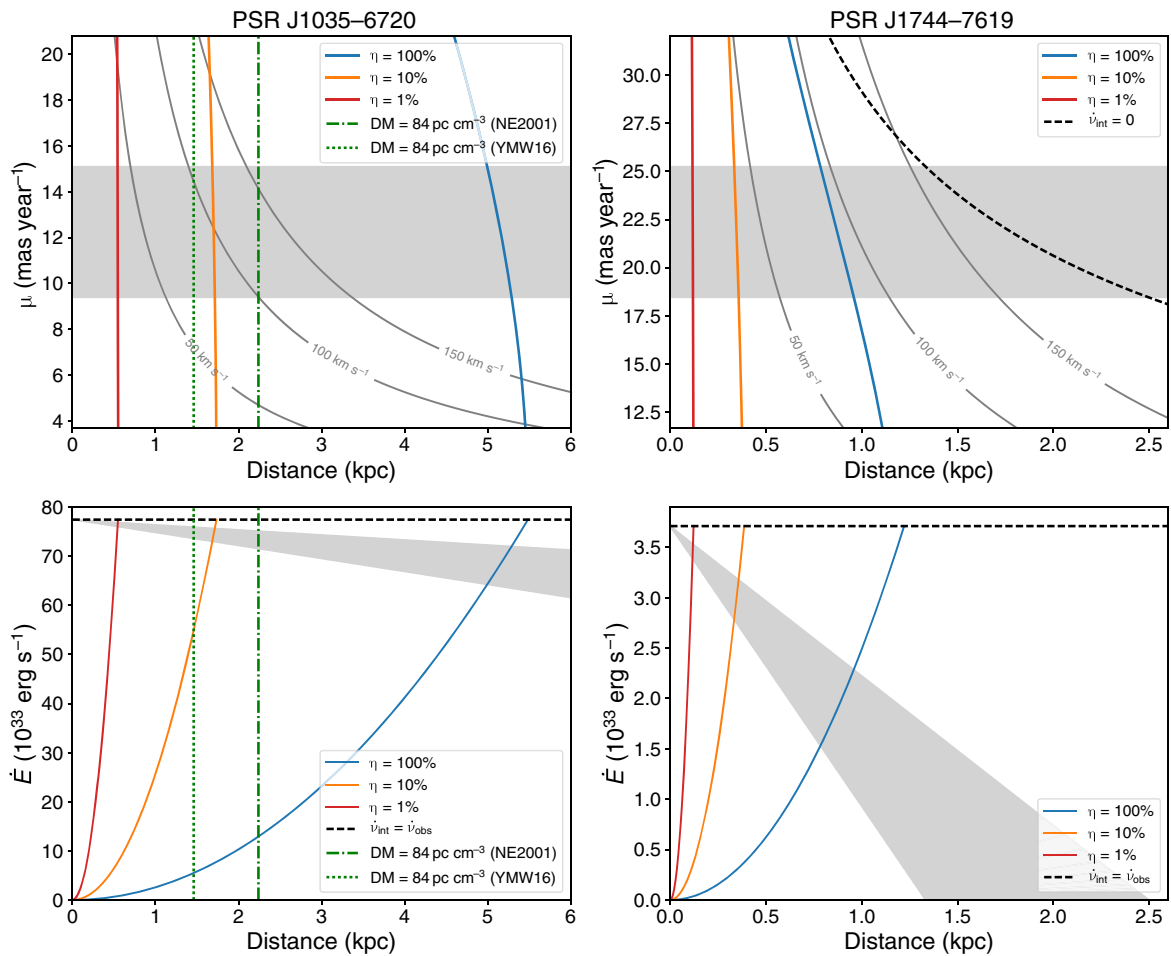

**Fig. 4. Distance and spin-down constraints for the two new MSPs obtained from their gamma-ray flux, radio dispersion measure, and timing measurements of their proper motion.** (**Top**) constraints on MSP distances, as in figure 11 of the work by Abdo *et al.* (*9*). Assuming certain gamma-ray efficiencies, constraints on the pulsar distance are inferred from the gamma-ray flux and spin-down power, after correcting the observed ν̇ for the Shklovskii effect due to the measured proper motion (1σ range given by the gray shaded region). The physically realistic region is to the left of the 100% efficiency line ($L_γ = \dot{E}$), although higher apparent efficiencies are possible, depending on the beam correction factor. Contours of constant transverse velocity are shown in gray. For PSR J1035−6720, the green dotted and dashed-dotted vertical lines show the DM distance according to the YMW16 and NE2001 models, respectively. For PSR J1744−7619, the ν̇$_{INT}$ = 0 line shows the required distance for the observed spin down to be purely due to the Shklovskii effect, giving a hard upper limit. (**Bottom**) Spin-down powers after correcting ν̇ for the Shklovskii effect. The shaded region shows the allowed range for $\dot{E}$ at each distance, given the observed proper motion. The solid curves show $\dot{E}$ as a function of distance at fixed gamma-ray efficiencies. The dashed line shows the maximum $\dot{E}$, that is, if there was no proper motion.

Figure 4 shows the results of this, assuming a canonical moment of inertia $I = 10^{45}$ g cm², a gamma-ray beam correction factor of $f_Ω = 1$ (*49*), and a realistic range of gamma-ray efficiencies $0.01 < η < 1$ (converting loss of rotational energy into gamma-ray emission) (*9*). From the inferred distances, we can retrieve Shklovskii-corrected spin-down powers ($\dot{E}$) via $\dot{E} = F_γ 4π f_Ω d^2 / η$. These ranges are shown in the lower panels of Fig. 4.

From its DM distance, the gamma-ray efficiency of PSR J1035−6720 is around 7% (YMW16 model) to 17% (NE2001 model). For PSR J1744−7619, the Shklovskii effect can account for up to 65% of the apparent spin-down rate of PSR J1744−7619. The observed gamma-ray flux constrains the distance to PSR J1744−7619 to be less than ~1 kpc, assuming η = 1 and $f_Ω = 1$.

If the gamma-ray efficiency of PSR J1744−7619 is high, then its spin-down power could be as low as $1.5 \times 10^{33}$ erg s⁻¹, which would make this one of the least energetic gamma-ray MSPs and close to the empirical gamma-ray "death line," that is, the minimum spin-down power an MSP must have to emit gamma rays (*50*).

An additional comparison can be made between the transverse velocities of known isolated MSPs and the transverse velocities given by the proper motion measurements and distance constraints described above. Lines of constant transverse velocity are shown in Fig. 4 to illustrate this. The mean (median) transverse velocity of isolated MSPs (outside of globular clusters) in the ATNF Catalogue is 84 (59) km s⁻¹. For PSR J1035−6720, the DM distances correspond to transverse velocities of

84 ± 20 km s⁻¹ (YMW16) and 130 ± 30 km s⁻¹ (NE2001). For PSR J1744−7619, assuming 100% gamma-ray efficiency leads to a transverse velocity of around 90 km s⁻¹, suggesting that the 1-kpc distance limit derived above is likely to be realistic for this pulsar. However, larger distances (for example, due to a lower beam correction factor) cannot be excluded on the basis of the resulting transverse velocity alone.

### Radio observations

The radio pulsar survey performed by Camilo et al. (6) targeted 56 unidentified Fermi-LAT sources with the Parkes radio telescope and detected 11 MSPs, 10 of which were new discoveries. As part of this survey, we observed the LAT sources now known to be associated with the two MSPs multiple times between 2009 and 2012, at a center frequency of 1390 MHz using an analog filter-bank system. The total power from the central beam of the Parkes multibeam receiver (full width at half maximum = 14′), filtered through 512 frequency channels spanning a bandwidth of 256 MHz, was sampled every 125 μs and recorded for off-line analysis.

Typical integration times were around 1 hour, although they ranged between 41 and 136 min for these two sources [see table 1 of Abdo et al. (7)]. The LAT source associated with PSR J1035−6720 was observed a total of nine times, at a typical offset from the actual pulsar position of 4.3′. This reduced the sensitivity of those observations to 80% of the maximum (on axis) sensitivity. The LAT source associated with PSR J1744−7619 was observed 10 times at a typical offset of 1′, with no material impact on the sensitivity.

The data were analyzed using standard pulsar search techniques implemented in PRESTO (51). Before the new gamma-ray pulsars were known, the data were searched as described by Camilo et al. (6), considering possible DMs up to approximately twice the maximum DM predicted by the NE2001 model (25) for the corresponding line of sight [about 500 pc cm⁻³ for PSR J1035−6720 and 200 pc cm⁻³ for PSR J1744−7619; table 1 of Abdo et al. (7)]. Subsequently, we reanalyzed the data sets using the timing ephemerides obtained from gamma-ray data, folding the radio data while searching only in DM. None of the observations yielded significant radio pulsations.

The nominal threshold for these searches depends on the putative radio pulse duty cycle of these pulsars and their DM. For an assumed duty cycle of 25% and DM < 50 pc cm⁻³, an indicative 1.4-GHz flux density threshold for PSR J1035−6720 is about 0.25 mJy normalized to a 1-hour integration. For PSR J1744−7619, the equivalent figure is approximately 0.15 mJy.

Some of the MSPs discovered by Camilo et al. (6) were not detected in all their search observations. This can be caused by a combination of orbital acceleration, eclipses, or interstellar scintillation. The first two effects are not relevant for the two new isolated MSPs. Depending on the distance of the new MSPs, especially if they are nearby objects, interstellar scintillation could very much modulate any putative radio flux density recorded by our Parkes observing system. In extreme cases, such as PSR J1514−4946 (6), this can lead to radio-loud MSPs being discoverable for less than one-third of the time. However, we know of no established isolated radio-loud MSP in the Galactic disk that is undetectable as much as 90% of the time.

Following the discoveries of the pulsars, we performed five long observations toward PSR J1035−6270 with the Parkes radio telescope. All observations were performed at a center frequency of 1369 MHz and sampled voltages were processed online with a digital polyphase filter bank into 2048 channels across a bandwidth of 256 MHz, for an effective sampling time of 4 μs. The filter-bank outputs were detected and folded online into 256-bin profiles using the known spin ephemeris. Before each observation, we observed a pulsed noise diode to facilitate polarization and flux density (via archival observation of the radio galaxy Hydra A) calibration. The first four observations used the wideband "H-OH" receiver, whereas the latter used the center pixel of the "multibeam" receiver; for consistency, we report results only from the first four observations.

We searched the first observation over a range of DMs and found a promising signal with a signal-to-noise ratio of ~ 10 and a duty cycle of about 10% at a DM of 84.2 pc cm⁻³. The signal appears at the same DM and pulse phase in the observations on 24 August 2016 and, particularly, 26 September 2016, confirming the detection. Using a Gaussian template, we maximized the likelihood for the data from 09 June 2016, 24 August 2016, and 26 September 2016 to estimate an optimal DM of 84.16 ± 0.22 pc cm⁻³. With this template and DM, we measured the average flux density in each observation. The results are summarized in Table 2. (The pulsed emission in the observation on 07 August 2016 is too weak to be detected independently, but with knowledge of the phase and profile shape, the flux density can be estimated.)

For PSR J1744−7619, we performed two long observations using an identical observational setup and the multibeam system. We performed a similar search over DM (up to 800 pc cm⁻³) and found no pulsed emission. Restricting consideration to DM < 200 pc cm⁻³, we smoothed the fully averaged profile at each trial DM with a top hat of width 12.5% and recorded the maximum flux density. The results depend modestly on the baseline subtraction method used, and taking the average of two methods gives 95% confidence flux density upper limits of 0.031 and 0.032 mJy, respectively, for the two observations and 0.023 mJy for their coadded sum.

Combined with our estimated distance upper limit of d < 1 kpc, we found an upper limit on the pseudo-luminosity of PSR J1744−7619 of $L_{1400} < 0.023$ mJy kpc². Only two known MSPs with both reliable flux density measurements and DM-independent distance estimates have lower pseudo-luminosities, PSR J1400−1431 with $L_{1400} = 0.01$ mJy kpc² (28) and PSR J2322−2650 with $L_{1400} = 0.008$ mJy kpc² (29). Of the 164 remaining MSPs in the ATNF Catalogue with reported flux density measurements and distance estimates (including those inferred from DM measurements), only two have apparent pseudo-luminosities below that of PSR J1744−7619. For one of these, PSR J0922−52, the reported flux density is only a lower bound because of an unknown pointing offset in its discovery observation (52). For the other, PSR J1745−0952, the YMW16 distance estimate (0.23 kpc) strongly disagrees with that from NE2001 (1.8 kpc). This pulsar lies toward the Galactic Loop I bubble, within which the YMW16 distance for the closest pulsar with an independently estimated distance (PSR J1744−1134, angular separation $\Delta\theta = 1.7°$ from PSR J1745−0952) was seen to be underestimated by a factor of 2.6 [parallax distance $d = 0.395$ kpc and DM distance $d_{DM} = 0.148$ kpc; (24)]. In combination with the extremely low apparent pseudo-luminosity, this suggests that the DM distance estimate is unreliable in this case.

**Table 2. Summary of dedicated follow-up radio observations with the Parkes radio telescope.**

| Target | Date | Duration (min) | Receiver | $S_{1400}$ (mJy) |
|---|---|---|---|---|
| PSR J1035−6720 | 09 June 2016 | 188 | H-OH | 0.052 ± 0.0064 |
| | 07 August 2016 | 164 | H-OH | 0.023 ± 0.0064 |
| | 24 August 2016 | 169 | H-OH | 0.039 ± 0.0061 |
| | 26 September 2016 | 232 | H-OH | 0.045 ± 0.005 |
| | 19 March 2017 | 180 | Multibeam | — |
| PSR J1744−7619 | 19 March 2017 | 180 | Multibeam | < 0.031 |
| | 10 April 2017 | 163 | Multibeam | < 0.032 |

## X-ray observations

The fields of PSRs J1035−6720 and J1744−7619 were observed by XMM-Newton for 25 ks (obsids 692830101 and 692830201) with the aim of detecting x-ray counterparts to possible radio-quiet MSPs in unidentified LAT sources (21). These gamma-ray sources were selected before the pulsation discoveries, because their gamma-ray properties indicated a probable pulsar nature.

Plausible x-ray counterparts for both pulsars were detected with a significance greater than $10\sigma$ at locations consistent with the newly discovered pulsars' timing positions. Their x-ray unabsorbed flux in the 0.3- to 10-keV energy range is $3.06^{+0.96}_{-0.50} \times 10^{-14}$ erg cm$^{-2}$ s$^{-1}$ for PSR J1035−6720 and $1.92^{+0.59}_{-0.39} \times 10^{-14}$ erg cm$^{-2}$ s$^{-1}$ for PSR J1744−7619. See table 11 of Saz Parkinson et al. (21) for additional information. We computed the probability that the association between the x-ray source and the MSP is due to a chance coincidence using $P = 1 - \exp(\pi\rho r^2)$, where $r$ is the matching radius (in our case, the x-ray source error radius) and $\rho$ is the density of x-ray objects in the XMM−Newton EPIC (European Photon Imaging Camera) field, regardless of their flux. We estimated $P \sim 1.3 \times 10^{-4}$ for PSR J1035−6720 ($r = 2''$ and $\rho \sim 0.038$ arc min$^{-2}$) and $P \sim 1.4 \times 10^{-4}$ for PSR J1744−7619 ($r = 1.9''$ and $\rho \sim 0.046$ arc min$^{-2}$), which make a chance positional coincidence unlikely. In addition, the gamma-ray–to–x-ray flux ratios of the two likely x-ray counterparts (~700 and ~1100) were consistent with an MSP nature (30), confirming their associations with PSRs J1035−6720 and J1744−7619.

## Gamma-ray pulse profile modeling

To model the gamma-ray emission geometry of the MSPs, we fit simulated pulse profiles to the observed photon phases, using the fitting technique described by Johnson et al. (31). We considered three emission models: an OG model, a TPC model, and a PSPC model. These are also described in the study of Johnson et al. (31) and references therein, and briefly summarized here. In the first two models, particle acceleration takes place in narrow "gaps" in the magnetosphere, where the plasma charge density deviates from the force-free configuration. In both models, these gaps border the last-closed magnetic field lines. In the OG model, the lower bound of the gap is defined by the "null-charge surface," where the plasma charge density changes sign. In the TPC model, the gap begins at the pulsar surface and extends to the light cylinder. The PSPC model is valid for low-$\dot{E}$ pulsars, where pair creation may be insufficient to reach the force-free configuration, allowing for particle acceleration throughout the regions of open field lines. We assumed a hollow-cone model (16) for the radio beams.

We used simulated rotation periods of 2.5 and 4.5 ms, respectively, and a simulated period derivative of $1 \times 10^{-20}$ s s$^{-1}$ for both. Instead of a Poisson likelihood, we used a $\chi^2$ statistic to fit the weighted-counts light curves, using 60 bins for both MSPs, with background levels and uncertainties calculated as in the study of Abdo et al. (9). The best-fit parameters (the angle between spin axis and magnetic dipole, $\alpha$; the angle between the spin axis and the line-of-sight, $\zeta$; and the estimated beam correction factor, $f_\Omega$) for each model are given in Table 3, with 95% confidence-level uncertainties estimated as described by Johnson et al.

(31). The best-fitting light curves from all models for each pulsar are shown in Fig. 5. It should be noted that, as pointed out by Pierbattista et al. (53), fitting only the gamma-ray light curves with these toy models can lead to systematic biases in the best-fit parameters of ~ 10°. We also note that no model considered here successfully reproduces the broad double-peaked pulse profile of PSR J1035−6720.

**Table 3. Best-fit parameters from gamma-ray pulse profile modeling.** For each pulsar, we report the best-fitting magnetic inclination angles ($\alpha$), viewing angles ($\zeta$), and beam correction factors ($f_\Omega$), according to the OG, TPC, and PSPC models.

| Parameter | PSR J1035−6720 | | | PSR J1744−7619 | | |
|---|---|---|---|---|---|---|
| Model | OG | TPC | PSPC | OG | TPC | PSPC |
| $-\ln(L)$ | 84.5 | 78.9 | 72.0 | 84.6 | 58.1 | 72.7 |
| $\alpha$ (deg) | $9^{+1}_{-2}$ | $7^{+2}_{-1}$ | $51^{+4}_{-3}$ | $67^{+2}_{-2}$ | $62^{+1}_{-3}$ | $62^{+5}_{-1}$ |
| $\zeta$ (deg) | $75^{+1}_{-1}$ | $70^{+1}_{-1}$ | $78^{+3}_{-3}$ | $32^{+2}_{-2}$ | $44^{+3}_{-3}$ | $74^{+5}_{-5}$ |
| $f_\Omega$ | $0.22^{+0.25}_{-0.04}$ | $0.62^{+0.08}_{-0.11}$ | $0.98^{+0.12}_{-0.04}$ | $0.94^{+0.05}_{-0.10}$ | $0.89^{+0.03}_{-0.09}$ | $1.14^{+0.15}_{-0.09}$ |

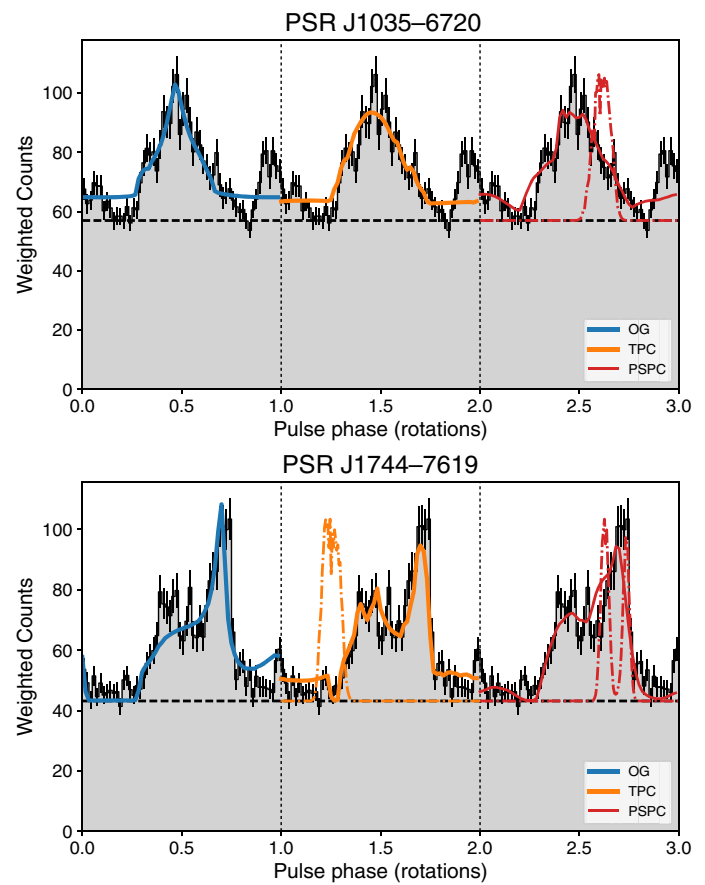

**Fig. 5. Gamma-ray pulse profiles of the newly detected MSPs.** The overlaying curves are the best-fitting pulse profiles predicted by fits to OG, TPC, and PSPC gamma-ray emission models. The black dashed line is the estimated background level, derived from the photon weights as in the study of Abdo et al. (9). Predicted radio pulse profiles (with arbitrary baseline and normalization) are shown by dashed-dotted lines. The three models are shown in separate rotations for clarity.

The PSPC model for PSR J1035−6720 predicted a phase offset between the gamma-ray and radio pulses that was reasonably consistent with that observed. However, the uncertainty in DM led to an additional uncertainty in the radio pulse phase of 0.15 rotations, precluding a joint fit of the radio and gamma-ray pulse profiles as is normally possible for other gamma-ray MSPs.

For the models that did not predict radio emission (TPC and OG for PSR J1035−6720 and OG for PSR J1744−7619), the lines of sight barely missed the edges of the emission cones at 300 MHz, suggesting that these MSPs might appear brighter at even lower frequencies.

**Acknowledgments:** We are grateful to the thousands of volunteers who have supported Einstein@Home. We also thank those volunteers whose computers discovered the new MSPs: PSR J1035–6720: "WSyS"; K. Kovacs, of Seattle, WA, USA, and the ATLAS Cluster, AEI, Hannover, Germany; PSR J1744–7619: D. Hoberer, of Gainesville, TX, USA, and the ATLAS Cluster, AEI, Hannover, Germany; I. Yakushin of Chicago, IL, USA, and the LIGO Laboratory, USA; and K. Pickstone of Oldham, UK. **Funding:** This work was supported by the Max-Planck-Gesellschaft, the Deutsche Forschungsgemeinschaft through an Emmy Noether Research Grant (no. PL 710/1-1 (H.J.P.)], and NSF award 1104902. C.J.C. acknowledges support from the European Research Council under the European Union's Horizon 2020 research and innovation programme (grant agreement no. 715051; Spiders). W.M. was partially supported by the Thailand Research Fund (grants TRG5880173 and RTA5980003). M.K. was funded by the Italian Ministry of Education, University and Research (contract no. FIRB-2012-RBFR12PM1F). Work at the Naval Research Laboratory is supported by NASA. The Parkes Observatory is part of the Australia Telescope, funded by the Commonwealth of Australia for operation as a National Facility managed by the Commonwealth Scientific and Industrial Research Organisation (CSIRO). The *Fermi* LAT Collaboration acknowledges generous ongoing support from a number of agencies and institutes that have supported both the development and the operation of the LAT, as well as scientific data analysis. These include the National Aeronautics and Space Administration and the Department of Energy in the United States, the Commissariat à l'Energie Atomique and the Centre National de la Recherche Scientifique/Institut National de Physique Nucléaire et de Physique des Particules in France; the Agenzia Spaziale Italiana and the Istituto Nazionale di Fisica Nucleare in Italy, the Ministry of Education, Culture, Sports, Science and Technology (MEXT), High Energy Accelerator Research Organization (KEK), and Japan Aerospace Exploration Agency (JAXA) in Japan; and the K. A. Wallenberg Foundation, the Swedish Research Council, and the Swedish National Space Board in Sweden. Additional support for science analysis during the operations phase is gratefully acknowledged from the Istituto Nazionale di Astrofisica in Italy and the Centre National d'Études Spatiales in France. This work performed in part under DOE Contract DE-AC02-76SF00515. **Author contributions:** C.J.C., H.J.P., J.W., and L.G. designed and implemented the Einstein@Home gamma-ray pulsar searches. M. Kerr performed follow-up searches with the Parkes radio telescope. T.J.J. carried out the gamma-ray pulse profile modeling. F.C. reanalyzed archival Parkes observations. D.S. performed analyses of x-ray observations. The manuscript was written by the aforementioned authors. The remaining authors contributed to the development and operation of the Einstein@Home project, to the design and operation of the *Fermi* LAT and analysis of its observations, or to the Fermi Pulsar Search Consortium. **Competing interests:** The authors declare that they have no competing interests. **Data and materials availability:** Fermi-LAT data are available from the Fermi Science Support Center (http://fermi.gsfc.nasa.gov/ssc). XMM-Newton data are available from the XMM-Newton Science Archive (http://nxsa.esac.esa.int). Parkes radio telescope data recorded for this paper will be made available from the CSIRO Data Access Portal (https://data.csiro.au/) 18 months after the observation dates and are available upon request before that. All data needed to evaluate the conclusions in the paper are present in the paper and/or the Supplementary Materials. Additional data related to this paper may be requested from the authors.

Submitted 18 August 2017
Accepted 24 January 2018
Published 28 February 2018
10.1126/sciadv.aao7228

# Science Advances

## Einstein@Home discovers a radio-quiet gamma-ray millisecond pulsar

Colin J. Clark, Holger J. Pletsch, Jason Wu, Lucas Guillemot, Matthew Kerr, Tyrel J. Johnson, Fernando Camilo, David Salvetti, Bruce Allen, David Anderson, Carsten Aulbert, Christian Beer, Oliver Bock, Andres Cuéllar, Heinz-Bernd Eggenstein, Henning Fehrmann, Michael Kramer, Shawn A. Kwang, Bernd Machenschalk, Lars Nieder, Markus Ackermann, Marco Ajello, Luca Baldini, Jean Ballet, Guido Barbiellini, Denis Bastieri, Ronaldo Bellazzini, Elisabetta Bissaldi, Roger D. Blandford, Elliott D. Bloom, Raffaella Bonino, Eugenio Bottacini, Terri J. Brandt, Johan Bregeon, Philippe Bruel, Rolf Buehler, Toby H. Burnett, Sara Buson, Rob A. Cameron, Regina Caputo, Patrizia A. Caraveo, Elisabetta Cavazzuti, Claudia Cecchi, Eric Charles, Alexandre Chekhtman, Stefano Ciprini, Lynn R. Cominsky, Denise Costantin, Sara Cutini, Filippo D'Ammando, Andrea De Luca, Rachele Desiante, Leonardo Di Venere, Mattia Di Mauro, Niccolò Di Lalla, Seth W. Digel, Cecilia Favuzzi, Elizabeth C. Ferrara, Anna Franckowiak, Yasushi Fukazawa, Stefan Funk, Piergiorgio Fusco, Fabio Gargano, Dario Gasparrini, Nico Giglietto, Francesco Giordano, Marcello Giroletti, Germán A. Gomez-Vargas, David Green, Isabelle A. Grenier, Sylvain Guiriec, Alice K. Harding, John W. Hewitt, Deirdre Horan, Guðlaugur Jóhannesson, Shiki Kensei, Michael Kuss, Giovanni La Mura, Stefan Larsson, Luca Latronico, Jian Li, Francesco Longo, Francesco Loparco, Michael N. Lovellette, Pasquale Lubrano, Jeffrey D. Magill, Simone Maldera, Alberto Manfreda, Mario N. Mazziotta, Julie E. McEnery, Peter F. Michelson, Nestor Mirabal, Warit Mitthumsiri, Tsunefumi Mizuno, Maria Elena Monzani, Aldo Morselli, Igor V. Moskalenko, Eric Nuss, Takashi Ohsugi, Nicola Omodei, Monica Orienti, Elena Orlando, Michele Palatiello, Vaidehi S. Paliya, Francesco de Palma, David Paneque, Jeremy S. Perkins, Massimo Persic, Melissa Pesce-Rollins, Troy A. Porter, Giacomo Principe, Silvia Rainò, Riccardo Rando, Paul S. Ray, Massimiliano Razzano, Anita Reimer, Olaf Reimer, Roger W. Romani, Pablo M. Saz Parkinson, Carmelo Sgrò, Eric J. Siskind, David A. Smith, Francesca Spada, Gloria Spandre, Paolo Spinelli, Jana B. Thayer, David J. Thompson, Diego F. Torres, Eleonora Troja, Giacomo Vianello, Kent Wood and Matthew Wood

| | |
|---|---|
| **ARTICLE TOOLS** | http://advances.sciencemag.org/content/4/2/eaao7228 |
| **REFERENCES** | This article cites 47 articles, 5 of which you can access for free<br>http://advances.sciencemag.org/content/4/2/eaao7228#BIBL |
| **PERMISSIONS** | http://www.sciencemag.org/help/reprints-and-permissions |

Use of this article is subject to the Terms of Service